\documentclass[prd,nofootinbib,preprint,floatfix]{revtex4}  
\usepackage[plainpages=false, colorlinks=true, anchorcolor=blue, linkcolor=blue, citecolor=blue, bookmarks=false]{hyperref}
\usepackage{url}
\usepackage{float}
\usepackage{siunitx}
\usepackage{multirow}

\usepackage{graphicx}
\newcommand{\rthis}[1]{\textcolor{black}{#1}}

\begin{document}
\author{Ambica  \surname{Govind} }%
 \altaffiliation{ep21btech11007@iith.ac.in}
\author{Shantanu  \surname{Desai}}%
 \altaffiliation{shntn05@gmail.com}
 
\affiliation{Department of Physics, Indian Institute of Technology, Hyderabad, Kandi, Telangana-502284  India}

\title{A test of MOND and Emergent Gravity with  SMACS J0723.3-7327 using eROSITA observations}

\begin{abstract}

 We implement a test of MOND and Verlinde's Emergent Gravity using the galaxy cluster SMACS J0723-7327, which has been recently  imaged using the  eROSITA X-ray telescope as well as with JWST. We test MOND using two independent methods. The first method involves comparing the dynamical MOND mass and baryonic mass, while the second method entails a comparison of the MOND-estimated temperature with the observed temperature.  We then  compare the unseen mass predicted by Emergent Gravity with the estimated dark matter mass. We find that MOND shows a mass discrepancy  in the central regions at high significance levels. The observed temperature profile is in marginal disagreement with that in the MOND paradigm. However, the Emergent Gravity Theory agrees in accurately accounting for the dynamical mass in the inner regions within $1\sigma$. Our results are qualitatively consistent with the  earlier tests on other clusters.

\end{abstract}
\maketitle

\section{Introduction}
One of the key ingredients of the standard $\Lambda$CDM Concordance model of the universe~\citep{Ratra} 
is  Cold Dark matter, which constitutes about 25\% of the total mass-energy budget. This cold dark matter component is known to be non-baryonic and decouples from the  remaining  components of the universe, while moving at non-relativistic velocities~\citep{JKG,Hooper,Bosma}. Although, this $\Lambda$CDM model agrees remarkably well with observational  data at large-scales~\citep{Planck2018},  there is no laboratory evidence for any cold dark matter candidate, despite over three  decades of searches via underground dark matter laboratory searches~\citep{Merritt} or  indirect dark matter searches through annihilation products~\cite{Gaskins,Desai04} or accelerator based searches~\citep{LHC}. We still do not know answers to even rudimentary questions about the CDM particle as to whether it is a fermion or boson, and also on  whether it is a thermal or  a non-thermal relic. Therefore, one alternative to obviate the need for dark matter is to construct a modified gravity theory,  which agrees with solar system and strong field tests of general relativity, but deviates in the ultra-weak field limit in dark matter dominated regions. Two such alternatives, which dispense with the need for dark matter include the modified Newtonian Dynamics (MOND) paradigm~\citep{Milgrom83} and Emergent gravity~\cite{verlinde2017emergent}, which is the subject of this paper.

One can think of MOND as a theoretical extension of Newtonian dynamics in the  low acceleration limit, i.e. for acceleration below some constant value $a_{0} \approx 10^{-10} \rm{m s^{-2}}$, where $a_{0}$ is    referred to  as Milgrom's constant. Milgrom  established a smooth transition between the dark matter dominated and the Newtonian regimes i.e. $g >> a_{0}$ and $g << a_{0}$ using the formula
\begin{equation} \label{eq1}
{\mathrm{\mu \left( \frac{g}{a_{0}} \right) \textbf{g} = \textbf{g\textsubscript{N}}}} ,
\end{equation}
where the interpolation function ($\mu(x)$) is given by, ${\mathrm{\mu \left( x \right) \rightarrow 1 \ for \ x \gg 1 \ and \ \mu \left( x \right) \rightarrow x \ for \ x\ll1}}$. Many examples of such MOND interpolation functions can be found in ~\citet{Famaey}.
The MOND paradigm  relates the observed gravitational acceleration $g$ and the Newtonian acceleration calculated from only the baryonic mass ($g_{bar}$), i.e. for acceleration above $a_{0}$, $g_{obs}\approx g_{bar}$ and for acceleration below the value, $g_{obs} \propto g_{bar}^{1/2}$ \citep{Famaey}. 

The MOND paradigm can also  explain some of the regularities and deterministic scaling relations at galactic scales such as the Radial acceleration relation (RAR)~\cite{RAR}, constancy of halo dark matter surface density~\cite{Donato}, baryonic Tully-Fisher relation~\cite{Famaey}. These successes cannot be trivially reproduced by the standard $\Lambda$CDM model, although some of the latest $\Lambda$CDM simulations can reproduce the observed RAR~\cite{PS21} and the constancy of dark matter surface density~\cite{Gopika23}. All the empirical successes of MOND which cannot be explained within $\Lambda$CDM have recently been reviewed in ~\cite{Banik}. Of course, there are also some tensions of MOND with observations of our own galaxy, where it is difficult to predict the observed decline in rotation curve at large distances~\cite{Chan23}. The RAR also does not seem to hold for elliptical galaxies~\cite{ChanDesai22}.

However, it has been known since a long time, that MOND (in its original form) does not work well for  relaxed isolated galaxy clusters in hydrostatic equilibrium~\citep{He88,Gerbel,Aguirre,Sanders,Angus,Silk,Seeram}. For relaxed galaxy clusters, MOND has been tested using two different methods in literature.
In the first of these methods, the  dynamical MOND masses for clusters  have been calculated using a modified equation of hydrostatic equilibrium and found to be much larger than the baryonic mass, consisting of hot diffuse gas and stars~\citep{He88,Sanders,Silk,Angus}. For the first ever cluster test of MOND using this method, which was   done for the Coma cluster ($z=0.023$), $a_0$ was kept as a free parameter and the estimated  value of $a_0$ was found to be roughly four times larger than that inferred for spiral galaxies, with an uncertainty of 20\%~\cite{He88}. Since the MOND mass scales with temperature according to $M_{mond} \propto T/a_0$, this can also be interpreted as the MOND dynamical mass to be much larger than the baryonic mass~\cite{Sanders}, although no statistical significance of the discrepancy could be ascertained. Soon thereafter, another test of MOND  using a  sample of eight Abell clusters upto redshift of $z \approx 0.06$ was carried out using data from the Einstein X-ray observatory~\cite{Gerbel}. The results from this analysis were somewhat ambiguous and inconclusive. The dynamical MOND mass was calculated in the asymptotic limit and then compared to the baryonic mass~\cite{Gerbel}.  The dynamical MOND mass was found to be greater than the baryonic mass in the central parts of the cluster, but comparable to the baryonic mass at large radii.
However, for some clusters, the MOND dynamical mass was found to be less than the total gas mass. Nevertheless, no error analysis was considered in this work~\cite{Gerbel}.
Then, ~\citet{Sanders} considered a sample of about 40 ROSAT-selected clusters~\cite{Reiprich} from $z=0.0556$ to $z=0.0936$ to test MOND. Using a double-$\beta$ profile, it was shown that the MOND dynamical mass is still greater than the baryonic mass, and the quoted ratio of the total excess dark residual  mass required within  MOND to the baryonic mass is equal to $1.6 \pm 1.7$~\cite{Sanders}.  However, despite the consistency within $1\sigma$, it was argued that MOND still required an additional dark matter mass, which is about ten times larger than the hot gas mass in the central core radii. This discrepancy could be alleviated using 2 eV neutrinos~\cite{Sanders}. Subsequently, ~\citet{Silk}
considered a sample of eight nearby clusters ($z \leq 0.15$) selected from XMM-Newton observations. They constructed an analytic expression for the ratio of MOND dynamical mass to Newtonian dynamical mass using the observed mass-temperature scaling relation and showed that it is equal to $1.63 \pm 0.29$, which amounts to  a $5\sigma$ discrepancy. All the aforementioned tests were done with clusters with masses of $\mathcal{O} (10^{14}-10^{15}) M_{\odot}$. The first test with clusters and galaxy groups for masses in the range   $10^{12}-10^{13} M_{\odot}$ was done in ~\cite{Angus}.  This work  considered a sample of 30 galaxy groups and low mass clusters from Chandra X-ray observations in the redshift range $z=0.009-0.081$~\cite{Gasta}.
For this analysis, a different MOND interpolation function was used to  relate  the dynamical MOND mass and Newtonian mass, as  compared to ~\cite{Sanders,Silk}. The errors in the MOND mass were estimated from the differences in two interpolating MOND functions as opposed to estimating them directly using the data. Once again, the dynamical mass was found to be greater than the Newtonian dynamical mass. After considering the contribution from 2 eV neutrinos, it was found that although neutrinos could explain the excess residual mass in the outer parts of groups/clusters for objects with $T>3$keV. However, they cannot explain the excess residual mass in the inner 150 Kpc for any of the groups/clusters hitherto considered. Although, no formal statistical significance was calculated, it was argued that no random or systematic errors could resolve the discrepancy. Therefore in all the three aforementioned works involving clusters and groups spanning a large range of redshifts and masses, it was shown that the dynamical MOND mass is  larger than the observed baryonic mass~\citep{Sanders,Silk,Angus}.

As opposed to comparing the dynamical modified mass with the baryonic mass, MOND has also been tested using the usual (Newtonian) equation of hydrostatic equilibrium, but plugging  the MONDian  formula for the acceleration, in order  to derive the resulting temperature profile~\citep{Aguirre}.  This work compared the MOND temperature for three nearby galaxy clusters (Virgo, Abell 2199, and  Coma), all located at $z \leq 0.03$  with the observed temperatures obtained from ROSAT X-ray observations. 
The  estimated temperature profile was found to be ten times smaller near  the cluster center ($r \leq 50 kpc$) than the observed temperature profiles, thus providing a strong challenge for MOND~\citep{Aguirre}. In addition to considering X-ray data for clusters in hydrostatic equilibrium, MOND has also been  been found to be inconsistent with weak and strong lensing data for clusters~\cite{Natarajan,Chiba}.

Therefore, the conclusion  is that  even   within the MOND paradigm, galaxy clusters  have unaccounted for missing mass. However, in all these works involving X-ray selected clusters, the objects used for analysis have been assumed to be in perfect hydrostatic equilibrium, with no accounting for the hydrostatic bias. This assumption is not always robust and the systematic errors due to hydrostatic bias could  be up to 20\%~\cite{massbias}. No such comparison or discussion of the errors due to hydrostatic bias has been  done in the aforementioned works. Therefore, it is important to redo these tests with new observations along with a  detailed characterization of the  error budget.

Recently, observational tests of some of the regularities found at galactic scales (which led to the MOND paradigm) such as the RAR and constancy of dark matter surface density have also been done on cluster scales. These results showed that the  acceleration scale and halo surface density were found to be  higher compared to galactic scales and different among cluster samples, implying that the RAR cannot be universal~\cite{Tian,Chan20,pradyumna2021yet,Pradyumna21,Eckert22,Chan22,Chan14,Gopika,Gopika21}.  Many relativistic theories of MOND are also in tension due to the  coincident gravitational wave and EM observations from GW 170817~\citep{Desai}.

In the last decade, a new modified theory of gravity has been proposed which combines the success of MOND at galactic scales with that of $\Lambda$CDM at cosmological scales~\cite{verlinde2017emergent}.
This model is known as Entropic or Emergent gravity. Emergent Gravity asserts that spacetime and gravity emerge together from the entanglement structure of an underlying microscopic theory. Vacuum excitations impart entropy content to space that manifests itself as dark energy. Baryonic matter displaces this dark energy and causes an elastic restoring force on matter-the additional 'dark gravity' hitherto posited to be the effect of dark matter in $\Lambda$CDM. A large number of tests of Emergent Gravity have been done using galaxy clusters~\cite{Ettori16,Ettori18,Miller,Nichol,ZuHone}, dwarf galaxies~\cite{Pardo}, dwarf spheroidal galaxies~\cite{Diez} and spiral galaxies~\cite{Lelli17,Yoon}, early-type galaxies~\cite{Tortora}, weak galaxy-galaxy lensing~\cite{Brouwer,Brouwer21,Luo21}.  The results from these works have provided mixed results for Emergent Gravity with some observations in agreement~\cite{Brouwer,Diez,Yoon,Brouwer21}, whereas others in tension~\cite{Lelli17,Pardo,Luo21}, while results with clusters  show both an agreement and disagreement depending on the distance from the cluster center~\cite{Ettori16,Ettori18,Miller,Nichol,ZuHone}.

In 2019, the eROSITA  satellite was launched, which will carry out an all-sky X-ray survey to discover new galaxy clusters and map out their properties with unprecedented precision and at the end of its survey will achieve a sensitivity of 25 times better than the ROSAT mission~\cite{Pre}. The eROSITA survey has already discovered a large number of galaxy clusters, and has done a detailed characterization of X-ray gas properties of these clusters including the error budget~\cite{Bulbul21}. 
In this work, we revisit some of the tests of MOND and Emergent Gravity previously carried out with relaxed cluster samples using one such galaxy cluster, which has been imaged with eROSITA and other telescopes in X-ray and optical, namely SMACS J0723.3-7327.

The manuscript is structured as follows. The dataset used for the analysis is described in Sect.~\ref{sec:dataset}. Our results for tests of MOND are outlined in Sect.~\ref{sec:analysis}. Our tests of Emergent  Gravity can be found in Sect.~\ref{sec:EG}.  We conclude in Sect.~\ref{sec:conclusions}.

\section{Dataset used for analysis}
\label{sec:dataset}
For this analysis, we use multi-wavelength observations of 
SMACS J0723.3-7327 (J0723 hereafter) described in ~\cite{liu2022x} (L22, hereafter). J0723 is a massive galaxy cluster located at a redshift of $z=0.39$ and with $M_{500}=(9.8 \pm 5.1) \times 10^{14} M_{\odot}$ and temperature given by $kT_{500} \sim 7$ keV.  This cluster has also been previously detected in X-Rays by ROSAT, Chandra and XMM-Newton,  in SZ by Planck and was the first lensing cluster  observed by the James Webb space telescope. The hydrostatic mass for J0723 is also in agreement with the lensing mass indicating that  there is no hydrostatic bias for this cluster~\cite{Caminha22}.
For this work we used X-ray observations of J0723 from eROSITA using their all-sky survey data from December 2019 to February 2022 with a total exposure time of 4.4~ks. It was scanned five times  during the first five eROSITA all-sky surveys.  In addition, archival Chandra observations  (from April 2014) were also used to constrain the temperature profile in the cluster center. More details on the observations and data reduction can be found in L22.  The temperature profiles were estimated by combining the Chandra and eROSITA data. The electron density profiles were provided separately for both  Chandra and eROSITA.  L22 find that the measurements of temperature, gas mass and density as well as the hydrostatic mass are consistent between Chandra and eROSITA within $R_{2500}$.
Along with the best-fit parameters for temperature and pressure,  $1\sigma$ error bars have also been provided for all the free parameters in L22. We note however the quoted uncertainties assume that all measurements are independent. To take into account the covariance, one would need to use the joint posteriors. However, these have not been made publicly available. Therefore, it is likely that our error estimates could be overestimated.

\section{TESTING MOND}
\label{sec:analysis}
We test MOND for J0723 using two different methods. In the first method we compare the dynamical mass in the MOND picture to the observationally inferred baryonic mass. In the second method, we compare the observed temperature profile to that predicted by MOND.
\subsection{Comparison of dynamical and observed masses}
\label{sec:MONDsanders}
In order to test MOND with this relaxed cluster, we follow the same  methodology as in ~\cite{Sanders}.
We assume the intracluster medium to be an isotropic sphere characterized by the equation of hydrostatic equilibrium (HSE):
\begin{equation}
    \frac{dP}{dr}=-\rho g,
    \label{eq:2}
\end{equation}
where  $P$ denotes the pressure, $\rho$ being the density, and  $g$ is the observed gravitational acceleration. Since the hydrostatic mass agrees with the lensing mass, the HSE assumption is justified.  To test MOND, we plug the value of $g$ from Eq.~\ref{eq1}.
Here, we use  $a_0=1.2 \times 10^{-8} cm/s^2$, which  is the critical acceleration that demarcates the MOND regime from the Newtonian regime. The MOND interpolation function we use here  is as follows~\cite{Famaey}:
\begin{equation}
    \mu(x)=\frac{x}{\sqrt{1+x^2}}
    \label{eq:3}
\end{equation}
We then obtain the   observed acceleration $g$ from Eq~\ref{eq:2} and assuming an ideal gas equation of state:
\begin{equation}
    g=-\frac{1}{\rho}\frac{dp}{dr}=-\frac{kT}{fm_pr} \left[ \frac{d \ln \rho}{d \ln r}+\frac{d \ln T}{d \ln r}\right] , 
    \label{eq:4}
\end{equation}
and the dynamical mass in the Newtonian regime is given by~\cite{Sarazin}:
\begin{equation}
    M_N=-\frac{kTr}{Gfm_p} \left[ \frac{d \ln \rho}{d \ln r}+\frac{d \ln T}{d \ln r}\right]
    \label{eq:5}
\end{equation}
If we further posit the interpolation function to be of the form as in Eq.~\ref{eq:3}, the dynamical mass in MOND regime comes out to be:
\begin{equation}
    M_m=\frac{M_N}{\sqrt{1+(\frac{a_0}{g})^2}}
  \label{eq:6}  
\end{equation}
In order to use Eq~\ref{eq:5}, we require the radial temperature and density profiles of the cluster. For this purpose we use the profiles provided in L22.
The  radial temperature profile of J0723 was fit to the following model originally used in~\cite{vikhlinin2006chandra} to model Chandra X-ray cluster sample:
\begin{equation}
    T(r)=T_0\frac{\left(r/r_{cool}\right)^{a_{cool}}+T_{min}/T_0}{\left(r/r_{cool}\right)^{a_{cool}}+1} \frac{\left(r/r_t\right)^{-a}}{\left[1+\left(r/r_t\right)^b\right]^{c/b}}
    \label{eq:7}
\end{equation}
L22 used the gas number density profiles also from ~\cite{vikhlinin2006chandra}, which is an augmented version of the $\beta$-profile~\cite{betamodel} but without the second $\beta$-component:
\begin{equation}
n_{\mathrm p}n_{\mathrm e} = n_0^2 \cdot \frac{(r/r_c)^{-\alpha}}{(1+r^2/r_c^2)^{3\beta-\alpha/2}}\frac{1}{(1+r^{\gamma}/{r_s}^{\gamma})^{\epsilon/\gamma}},
\label{eq:8}
\end{equation}
and the corresponding  gas  mass density is calculated using:
\begin{equation}
    \rho_{gas}=1.624 m_p \sqrt{n_p n_e}
    \label{eq:9}
\end{equation}

L22 finds that the above model can explain its observational profiles well, where the Markov Chain Monte Carlo  fits to the parameters can be found.

The above relations are employed to calculate the dynamical mass in the Newtonian and thereby in the MOND picture. 
We then calculate the baryonic mass ($M_{baryon}$), which is given by the sum of gas ($M_{gas}$) and star mass ($M_{\star}$).
We calculate the mass of the gas within a radius $r$  (assuming spherical symmetry) to be:
\begin{equation}
    M_{gas}(r)=4\pi \int_0^r r^2\rho_{gas}(r) dr
\end{equation}
Similar to our work on the tests of linearity of dark matter to baryonic mass~\cite{Varenya}, we used the  estimate of the stellar mass from the stellar-gas mass relations  derived in ~\cite{chiu2018baryon}:
\begin{equation}
\label{eq:mstar_mgas}
M_{\star}(r) = 4 \times 10^{12} M_{\odot} \times \left(\frac{M_{\text{gas}}(r)}{5.7 \times 10^{13} M_{\odot}}\right)^{0.60 \pm 0.10}
\end{equation} 
This relation was obtained  using a sample of 91 SPT-SZ clusters detected up to redshift of 1.3 and includes contributions from the brightest cluster galaxy as well as cluster member galaxies within $r_{500}$. We note however,  that this relation does not include the contribution from intracluster light (ICL), which is difficult to constrain at high redshifts~\cite{Contini}. However, because of the exquisite imaging of the JWST, the ICL has been detected in this cluster~\cite{Mahler,Pascale}, although no estimate of the mass is available at the time of writing. Although the ICL could be a significant component of the total stellar mass of the cluster, neglecting it will not affect our final analysis, since the stellar mass is much smaller than the gas mass. However, the empirical relation in Eq.~\ref{eq:mstar_mgas} includes the stellar mass contributions from the  BCG as well as cluster member galaxies within $r_{500}$~\cite{chiu2018baryon}. The BCG stellar mass for J0723-7327 has also been estimated using ground-based $Ks$ photometry and found to be equal to $3.6 \times 10^{11} M_{\odot}$, with an uncertainty of 20\%~\cite{Bellstedt}. It is however possible that the detailed  radial profile for the stellar mass within the BCG could be different as compared to the empirical model we have assumed. For example, a Hernquist profile was used to model the BCG stellar mass profile in ~\cite{Tian}. To get this dynamical profile, one would need to carry out detailed SED modelling of the inner BCG core using JWST data, which is beyond the scope of this work.

\begin{figure}[h!]
    \centering
    \includegraphics[width=100mm,height=80mm]{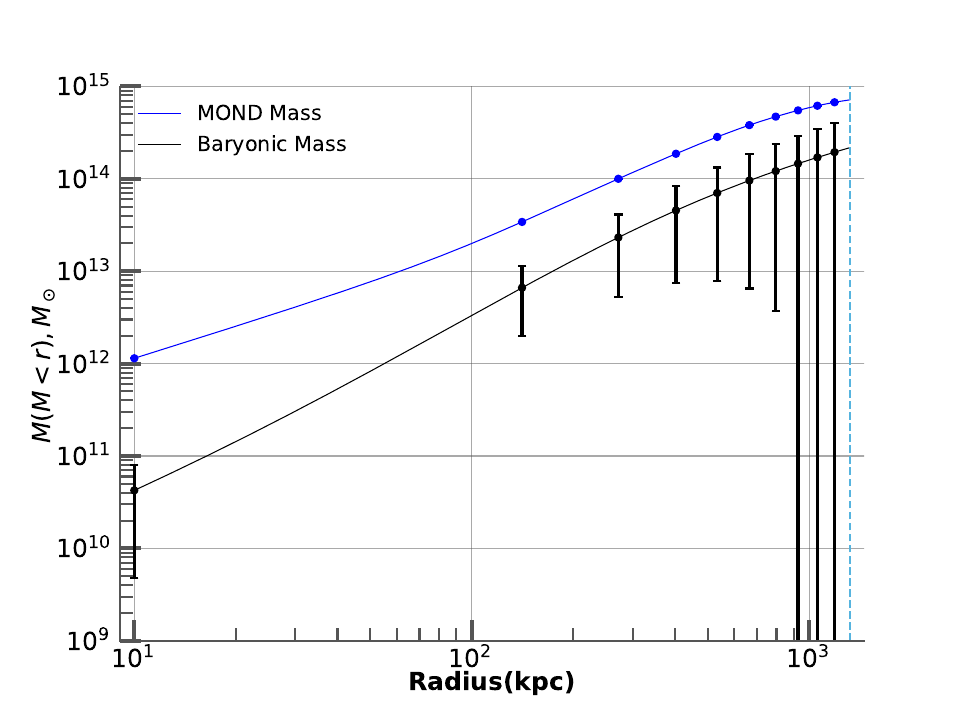}
    \caption{A comparison of the dynamical mass in the MOND and Newtonian regime with the baryonic mass observed from eROSITA  as a function of distance from the cluster center plotted until $r_{500}$. The $1\sigma$ error bars indicate the  uncertainties as calculated by error propagation. The uncertainty in the  baryonic mass is pronounced at large radii due to errors in $n_0$ and $r_c$. The data points are shown at 10 arbitrary radii equally spaced on the linear scale.}
    \label{fig:1}
\end{figure}

In order to test MOND, we compare the dynamical mass of the cluster within MOND with the baryonic mass \rthis{and quantify the statistical significance of the  discrepancy.}
We calculate both the baryonic and MOND dynamical masses at 1000 radii uniformly distributed between the cluster center and $r_{500}$. The errors in the  MOND and baryonic mass were obtained through error propagation, using the $1\sigma$ error bars provided in L22 for all the free parameters of the gas and temperature profiles.
This comparison of the MOND dynamical mass and observed baryonic mass can be found in Fig.~\ref{fig:1}, where we show masses corresponding to 10 radii. 
At low radii, the MOND masses are much higher in magnitude than the total baryonic mass, with a maximum disparity of $29 \sigma$. 
The disparity quickly reduces with the radius and at $r\approx 1000$kpc, the discrepancy reduces to around 2.6$\sigma$.
Therefore MOND performs better at larger radii, and there still remains a large amount of ``missing mass'' close to the cluster center at around 29$\sigma$, which is  unaccounted for by MOND.

\subsection{Temperature Profiles}
\begin{figure}[H]
    \centering
    \includegraphics[width=100mm,height=80mm]{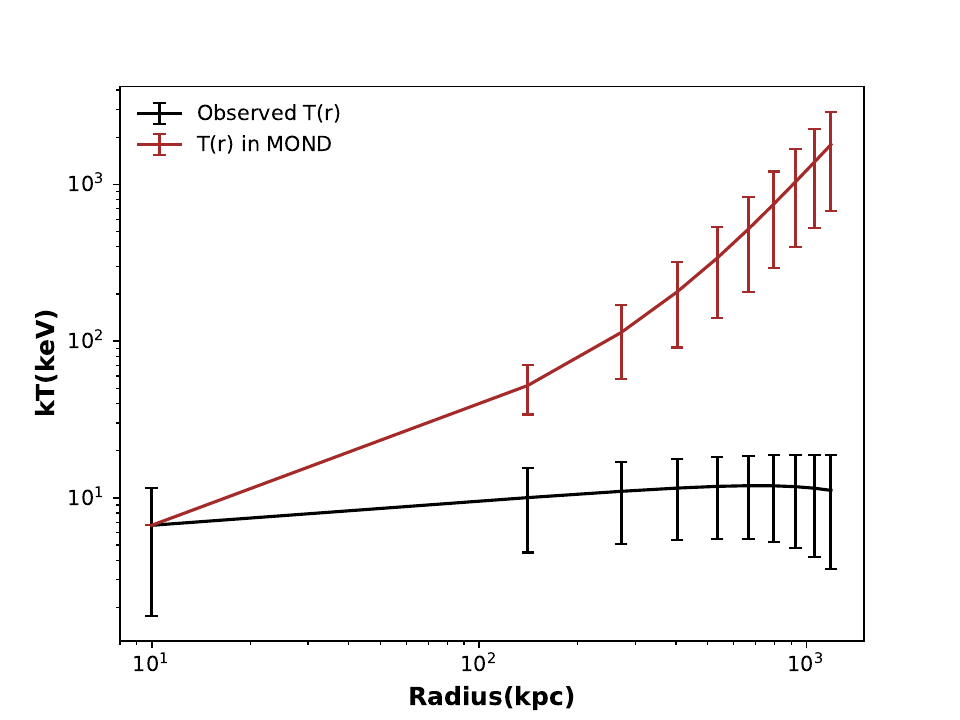}
    \caption{A comparison of the observed temperature profile against the numerically calculated temperature profile in the MOND framework plotted until $r_{500}$. The data points are shown at 10 arbitrary radii equally spaced on the linear scale. We estimated the errors in the numerically calculated temperature by generating 1000 normally distributed samples of the parameters about their value with a standard deviation equal to their errors, omitting samples with values outside these errors and then calculating the temperature extrema at each radius.}
    \label{fig:2}
\end{figure}

We now test MOND using  the same methodology as in ~\cite{Aguirre}. Here, instead of comparing the dynamical MOND and baryonic mass, we solve for  the temperature profile, by using the MONDian acceleration
in HSE,  and then compare that to the  observed temperature profile. We briefly recap this procedure.  In the limit $g<<a_0$, Eq.~\ref{eq:4} can be rearranged as:
\begin{equation}
    \frac{d \ln \rho}{d \ln r}+\frac{d \ln T}{d \ln r}= -\frac{gfm_pr}{kT}=-\frac{\sqrt{g_Na_0}fm_pr}{kT}=-\frac{\sqrt{GM_ma_0}fm_p}{kT}
    \label{eq:12}
\end{equation}
 To calculate the temperature profile in the MOND and hydrostatic equilibrium scenario, we numerically calculate $T(r)$ from Eq.~\ref{eq:12} given the observed $\rho(r)$ (where we used the eROSITA estimated parameters) and the dynamical mass in the MOND Scenario. The calculation is done via a fourth-order Runge-Kutta method with the initial condition that $T (10$ kpc$)$ in Eq.~\ref{eq:12} is equal to the observed temperature. The uncertainties  are calculated by generating 1000 normally distributed samples for each of the six free parameters of the density profile with their $1\sigma$ uncertainties as the standard deviation, excluding samples beyond the bounds set by the uncertainties and then computing the temperature extrema. Once again, we show the observed and MOND-estimated temperature at about 10 radii from cluster center until $r_{500}$. This plot can be seen in Fig.~\ref{fig:2}.
Although both the temperatures agree (by construction) at 10 kpc, we find that  MOND predicts a higher temperature than the observed temperature at all radii, with the disparity only increasing with $r$. 
We find a maximum difference of around $2.46\sigma$ at a radius of 84.6 kpc  and at $r_{500}$, the disagreement  is about  $1.6 \sigma$. To conclude, we find using this method that the observed and estimated temperature profiles only show a marginal disagreement, and it is not as pronounced as the analysis in Sect.~\ref{sec:MONDsanders}. 

\section{TESTING EMERGENT GRAVITY}
\label{sec:EG}
In order to test Emergent Gravity, we follow the same prescription as ~\cite{Ettori16},  which derived the  following relation between  the  baryonic  mass and the emergent ``dark  matter'',  which arises due to elastic response resulting from the entropy displacement:
\begin{equation}
    M_{DM,EG}^2(r)=\frac{cH_0}{6G}r^2M_{baryon}(r)(1+3\delta_B)
\end{equation}
where $\delta_B$ =$\rho_B(r)/\bar{\rho_B}$, with $\bar{\rho_B}=M_{baryon}(r)/V(<r)$ denoting  the mean baryon density within a spherical volume of radius $r$.
In our test,  we calculate the mass of the emergent dark matter component and compare it with the Newtonian estimate of dark matter given by $M_N-M_{baryon}$.
The Newtonian and baryonic mass are estimated in the same way as in Sect.~\ref{sec:MONDsanders}.
\begin{figure}[H]
    \centering
    \includegraphics[width=100mm,height=80mm]{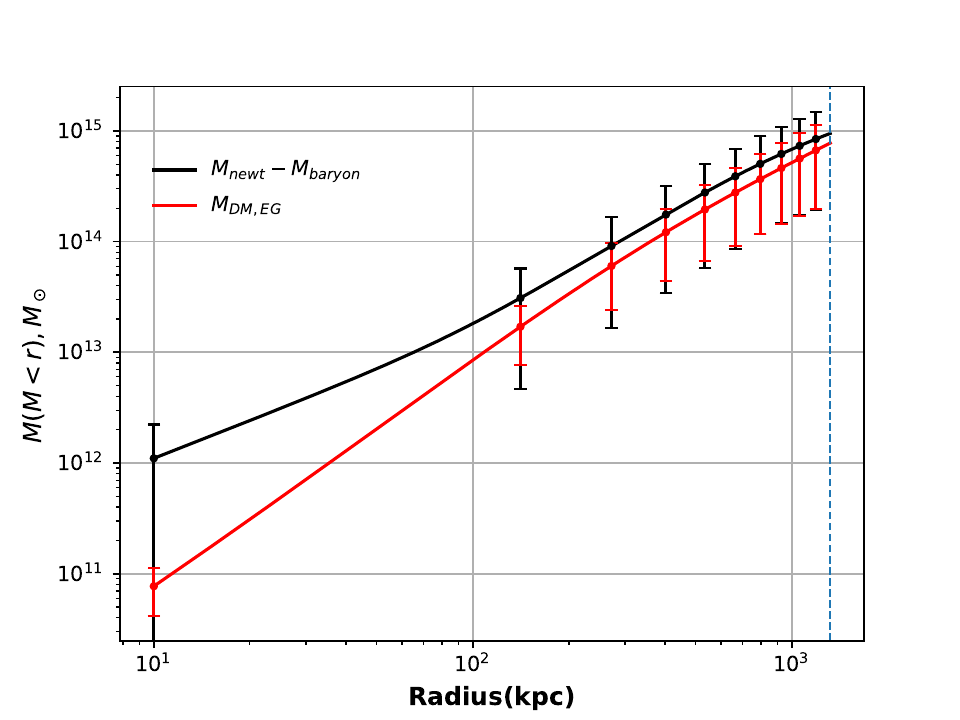}
    \caption{\small{A comparison of the mass deficit with the dark matter estimate in the Emergent Gravity formulation until $r_{500}$ indicated by the blue dotted line. The $1\sigma$ error bars show the  uncertainties calculated by error propagation. The mass deficit has a very high uncertainty at small radii as the error contribution from each of the 14 parameters adds up collectively.} }
    \label{fig:3}
\end{figure}
A comparison of the two can be found in Fig.~\ref{fig:3}.
From Fig~\ref{fig:3}, we find that emergent dark matter component in EG is consistent with the Newtonian dark matter mass to within 1$\sigma$ at all radii. Therefore, we find that the EG paradigm is  consistent with observations for SMACS J0723.3-7327.

\section{Conclusions}
\label{sec:conclusions}
In this work we have used X-Ray observations from eROSITA (discussed in L22) to test two modified gravity theories (namely MOND and Emergent Gravity) which dispense with the need for  dark matter, using the  cluster J0723 observed recently with eROSITA and JWST.   We tested MOND using two independent methods. In the first method, we compared the dynamical mass in MOND with the total baryonic mass. In the second method, we compared the MOND  inferred temperature profile with the observed temperature. In order to test  Emergent Gravity,
we compared  the excess mass predicted with Emergent Gravity with the observed Newtonian missing mass.
For all the above calculations, we  used temperature and density profiles provided in L22. Our conclusions are as follows:
\begin{itemize}
    \item The first test renders MOND dynamical mass to be discrepant  with respect to the baryonic mass far away from the cluster center ($\approx 1000$ kpc) at $2.6\sigma$. However,  close to the cluster center, we find that the largest discrepancy amounts  to  $29 \sigma$ at 10 kpc, which implies that MOND fails egregiously. 
    \item The second test implies that MOND predicts a temperature profile, which is marginally  higher  then the observed temperature at all radii, with a maximum discrepancy of $2.5\sigma$.
    \item The third test conveys that Emergent Gravity,  is consistent  with the observed mass deficit using this emergent dark matter at all radii to within $1\sigma$. This is  consistent with some of the earlier studies. 
\end{itemize}
We should however note that one caveat in our analysis is that we have not taken into consideration the covariances between the quoted parameter uncertainties, and hence it is likely that some of our errors could have been overestimated. Despite that, we still see large discrepancies in our MOND at small radii.
One way to reconcile the discrepancy of MOND  with observations only when there is evidence of additional missing baryons close to the cluster center,  or one needs to posit non-local versions of MOND~\cite{Woodard14,Woodard16} or some of the most recent relativistic versions proposed in ~\cite{Skordis}. \rthis{However, the fact that Emergent Gravity works for the same cluster is interesting. Nevertheless, more extensive tests are needed with additional clusters, since other tests with clusters   did not show an agreement with Emergent Gravity at all radii~\cite{Nichol,ZuHone,Ettori18}. More theoretical developments including solutions of  geodesic equations  within Emergent gravity are also needed in order to test this theory using cluster lensing data. Some progress along this direction has already been done such as developing a covariant formulation of Emergent Gravity~\cite{Sabine}.}

Efforts from various fields in the scientific community are underway to address these, in the form of precise observations, dark matter searches and theoretical development. On the observational side, we expect an improvement of results with better quality of X-Ray data powered by more sophisticated missions such as Athena~\cite{Athena} or eXTP~\cite{eXTP} to be deployed within the next decade. This data would render more precise tests of MOND and Emergent Gravity, among others.

\section*{Acknowledgements}
We are grateful to the anonymous referee for constructive feedback on our manuscript and to I-Non Chiu for very helpful  correspondence.

\bibliography{main}

\begin{thebibliography}{73}
\expandafter\ifx\csname natexlab\endcsname\relax\def\natexlab#1{#1}\fi
\expandafter\ifx\csname bibnamefont\endcsname\relax
  \def\bibnamefont#1{#1}\fi
\expandafter\ifx\csname bibfnamefont\endcsname\relax
  \def\bibfnamefont#1{#1}\fi
\expandafter\ifx\csname citenamefont\endcsname\relax
  \def\citenamefont#1{#1}\fi
\expandafter\ifx\csname url\endcsname\relax
  \def\url#1{\texttt{#1}}\fi
\expandafter\ifx\csname urlprefix\endcsname\relax\def\urlprefix{URL }\fi
\providecommand{\bibinfo}[2]{#2}
\providecommand{\eprint}[2][]{\url{#2}}

\bibitem[{\citenamefont{{Ratra} and {Vogeley}}(2008)}]{Ratra}
\bibinfo{author}{\bibfnamefont{B.}~\bibnamefont{{Ratra}}} \bibnamefont{and} \bibinfo{author}{\bibfnamefont{M.~S.} \bibnamefont{{Vogeley}}}, \bibinfo{journal}{\pasp} \textbf{\bibinfo{volume}{120}}, \bibinfo{pages}{235} (\bibinfo{year}{2008}), \eprint{0706.1565}.

\bibitem[{\citenamefont{{Jungman} et~al.}(1996)\citenamefont{{Jungman}, {Kamionkowski}, and {Griest}}}]{JKG}
\bibinfo{author}{\bibfnamefont{G.}~\bibnamefont{{Jungman}}}, \bibinfo{author}{\bibfnamefont{M.}~\bibnamefont{{Kamionkowski}}}, \bibnamefont{and} \bibinfo{author}{\bibfnamefont{K.}~\bibnamefont{{Griest}}}, \bibinfo{journal}{\physrep} \textbf{\bibinfo{volume}{267}}, \bibinfo{pages}{195} (\bibinfo{year}{1996}), \eprint{hep-ph/9506380}.

\bibitem[{\citenamefont{{Bertone} et~al.}(2005)\citenamefont{{Bertone}, {Hooper}, and {Silk}}}]{Hooper}
\bibinfo{author}{\bibfnamefont{G.}~\bibnamefont{{Bertone}}}, \bibinfo{author}{\bibfnamefont{D.}~\bibnamefont{{Hooper}}}, \bibnamefont{and} \bibinfo{author}{\bibfnamefont{J.}~\bibnamefont{{Silk}}}, \bibinfo{journal}{\physrep} \textbf{\bibinfo{volume}{405}}, \bibinfo{pages}{279} (\bibinfo{year}{2005}), \eprint{hep-ph/0404175}.

\bibitem[{\citenamefont{{Bosma}}(2023)}]{Bosma}
\bibinfo{author}{\bibfnamefont{A.}~\bibnamefont{{Bosma}}}, \bibinfo{journal}{arXiv e-prints} \bibinfo{eid}{arXiv:2309.06390} (\bibinfo{year}{2023}), \eprint{2309.06390}.

\bibitem[{\citenamefont{Ade et~al.}(2016)}]{Planck2018}
\bibinfo{author}{\bibfnamefont{P.}~\bibnamefont{Ade}} \bibnamefont{et~al.} (\bibinfo{collaboration}{Planck}), \bibinfo{journal}{Astron. Astrophys.} \textbf{\bibinfo{volume}{594}}, \bibinfo{pages}{A13} (\bibinfo{year}{2016}), \eprint{1502.01589}.

\bibitem[{\citenamefont{{Merritt}}(2017)}]{Merritt}
\bibinfo{author}{\bibfnamefont{D.}~\bibnamefont{{Merritt}}}, \bibinfo{journal}{Studies in the History and Philosophy of Modern Physics} \textbf{\bibinfo{volume}{57}}, \bibinfo{pages}{41} (\bibinfo{year}{2017}), \eprint{1703.02389}.

\bibitem[{\citenamefont{{Gaskins}}(2016)}]{Gaskins}
\bibinfo{author}{\bibfnamefont{J.~M.} \bibnamefont{{Gaskins}}}, \bibinfo{journal}{Contemporary Physics} \textbf{\bibinfo{volume}{57}}, \bibinfo{pages}{496} (\bibinfo{year}{2016}), \eprint{1604.00014}.

\bibitem[{\citenamefont{{Desai} et~al.}(2004)\citenamefont{{Desai}, {Ashie}, {Fukuda}, {Fukuda}, {Ishihara}, {Itow}, {Koshio}, {Minamino}, {Miura}, {Moriyama} et~al.}}]{Desai04}
\bibinfo{author}{\bibfnamefont{S.}~\bibnamefont{{Desai}}}, \bibinfo{author}{\bibfnamefont{Y.}~\bibnamefont{{Ashie}}}, \bibinfo{author}{\bibfnamefont{S.}~\bibnamefont{{Fukuda}}}, \bibinfo{author}{\bibfnamefont{Y.}~\bibnamefont{{Fukuda}}}, \bibinfo{author}{\bibfnamefont{K.}~\bibnamefont{{Ishihara}}}, \bibinfo{author}{\bibfnamefont{Y.}~\bibnamefont{{Itow}}}, \bibinfo{author}{\bibfnamefont{Y.}~\bibnamefont{{Koshio}}}, \bibinfo{author}{\bibfnamefont{A.}~\bibnamefont{{Minamino}}}, \bibinfo{author}{\bibfnamefont{M.}~\bibnamefont{{Miura}}}, \bibinfo{author}{\bibfnamefont{S.}~\bibnamefont{{Moriyama}}}, \bibnamefont{et~al.}, \bibinfo{journal}{\prd} \textbf{\bibinfo{volume}{70}}, \bibinfo{eid}{083523} (\bibinfo{year}{2004}), \eprint{hep-ex/0404025}.

\bibitem[{\citenamefont{Canepa}(2019)}]{LHC}
\bibinfo{author}{\bibfnamefont{A.}~\bibnamefont{Canepa}}, \bibinfo{journal}{Reviews in Physics} \textbf{\bibinfo{volume}{4}}, \bibinfo{pages}{100033} (\bibinfo{year}{2019}).

\bibitem[{\citenamefont{{Milgrom}}(1983)}]{Milgrom83}
\bibinfo{author}{\bibfnamefont{M.}~\bibnamefont{{Milgrom}}}, \bibinfo{journal}{\apj} \textbf{\bibinfo{volume}{270}}, \bibinfo{pages}{365} (\bibinfo{year}{1983}).

\bibitem[{\citenamefont{{Verlinde}}(2017)}]{verlinde2017emergent}
\bibinfo{author}{\bibfnamefont{E.~P.} \bibnamefont{{Verlinde}}}, \bibinfo{journal}{SciPost Physics} \textbf{\bibinfo{volume}{2}}, \bibinfo{eid}{016} (\bibinfo{year}{2017}), \eprint{1611.02269}.

\bibitem[{\citenamefont{{Famaey} and {McGaugh}}(2012)}]{Famaey}
\bibinfo{author}{\bibfnamefont{B.}~\bibnamefont{{Famaey}}} \bibnamefont{and} \bibinfo{author}{\bibfnamefont{S.~S.} \bibnamefont{{McGaugh}}}, \bibinfo{journal}{Living Reviews in Relativity} \textbf{\bibinfo{volume}{15}}, \bibinfo{eid}{10} (\bibinfo{year}{2012}), \eprint{1112.3960}.

\bibitem[{\citenamefont{{McGaugh} et~al.}(2016)\citenamefont{{McGaugh}, {Lelli}, and {Schombert}}}]{RAR}
\bibinfo{author}{\bibfnamefont{S.~S.} \bibnamefont{{McGaugh}}}, \bibinfo{author}{\bibfnamefont{F.}~\bibnamefont{{Lelli}}}, \bibnamefont{and} \bibinfo{author}{\bibfnamefont{J.~M.} \bibnamefont{{Schombert}}}, \bibinfo{journal}{\prl} \textbf{\bibinfo{volume}{117}}, \bibinfo{eid}{201101} (\bibinfo{year}{2016}), \eprint{1609.05917}.

\bibitem[{\citenamefont{{Donato} et~al.}(2009)\citenamefont{{Donato}, {Gentile}, {Salucci}, {Frigerio Martins}, {Wilkinson}, {Gilmore}, {Grebel}, {Koch}, and {Wyse}}}]{Donato}
\bibinfo{author}{\bibfnamefont{F.}~\bibnamefont{{Donato}}}, \bibinfo{author}{\bibfnamefont{G.}~\bibnamefont{{Gentile}}}, \bibinfo{author}{\bibfnamefont{P.}~\bibnamefont{{Salucci}}}, \bibinfo{author}{\bibfnamefont{C.}~\bibnamefont{{Frigerio Martins}}}, \bibinfo{author}{\bibfnamefont{M.~I.} \bibnamefont{{Wilkinson}}}, \bibinfo{author}{\bibfnamefont{G.}~\bibnamefont{{Gilmore}}}, \bibinfo{author}{\bibfnamefont{E.~K.} \bibnamefont{{Grebel}}}, \bibinfo{author}{\bibfnamefont{A.}~\bibnamefont{{Koch}}}, \bibnamefont{and} \bibinfo{author}{\bibfnamefont{R.}~\bibnamefont{{Wyse}}}, \bibinfo{journal}{\mnras} \textbf{\bibinfo{volume}{397}}, \bibinfo{pages}{1169} (\bibinfo{year}{2009}), \eprint{0904.4054}.

\bibitem[{\citenamefont{{Paranjape} and {Sheth}}(2021)}]{PS21}
\bibinfo{author}{\bibfnamefont{A.}~\bibnamefont{{Paranjape}}} \bibnamefont{and} \bibinfo{author}{\bibfnamefont{R.~K.} \bibnamefont{{Sheth}}}, \bibinfo{journal}{\mnras} \textbf{\bibinfo{volume}{507}}, \bibinfo{pages}{632} (\bibinfo{year}{2021}), \eprint{2102.13116}.

\bibitem[{\citenamefont{{Gopika} et~al.}(2023)\citenamefont{{Gopika}, {Desai}, and {Paranjape}}}]{Gopika23}
\bibinfo{author}{\bibfnamefont{K.}~\bibnamefont{{Gopika}}}, \bibinfo{author}{\bibfnamefont{S.}~\bibnamefont{{Desai}}}, \bibnamefont{and} \bibinfo{author}{\bibfnamefont{A.}~\bibnamefont{{Paranjape}}}, \bibinfo{journal}{\mnras} \textbf{\bibinfo{volume}{523}}, \bibinfo{pages}{1718} (\bibinfo{year}{2023}), \eprint{2303.12859}.

\bibitem[{\citenamefont{{Banik} and {Zhao}}(2022)}]{Banik}
\bibinfo{author}{\bibfnamefont{I.}~\bibnamefont{{Banik}}} \bibnamefont{and} \bibinfo{author}{\bibfnamefont{H.}~\bibnamefont{{Zhao}}}, \bibinfo{journal}{Symmetry} \textbf{\bibinfo{volume}{14}}, \bibinfo{pages}{1331} (\bibinfo{year}{2022}), \eprint{2110.06936}.

\bibitem[{\citenamefont{{Chan} and {Law}}(2023)}]{Chan23}
\bibinfo{author}{\bibfnamefont{M.~H.} \bibnamefont{{Chan}}} \bibnamefont{and} \bibinfo{author}{\bibfnamefont{K.~C.} \bibnamefont{{Law}}}, \bibinfo{journal}{\apj} \textbf{\bibinfo{volume}{957}}, \bibinfo{eid}{24} (\bibinfo{year}{2023}).

\bibitem[{\citenamefont{{Chan} et~al.}(2022)\citenamefont{{Chan}, {Desai}, and {Del Popolo}}}]{ChanDesai22}
\bibinfo{author}{\bibfnamefont{M.~H.} \bibnamefont{{Chan}}}, \bibinfo{author}{\bibfnamefont{S.}~\bibnamefont{{Desai}}}, \bibnamefont{and} \bibinfo{author}{\bibfnamefont{A.}~\bibnamefont{{Del Popolo}}}, \bibinfo{journal}{\pasj} \textbf{\bibinfo{volume}{74}}, \bibinfo{pages}{1441} (\bibinfo{year}{2022}), \eprint{2205.07515}.

\bibitem[{\citenamefont{{The} and {White}}(1988)}]{He88}
\bibinfo{author}{\bibfnamefont{L.~S.} \bibnamefont{{The}}} \bibnamefont{and} \bibinfo{author}{\bibfnamefont{S.~D.~M.} \bibnamefont{{White}}}, \bibinfo{journal}{\aj} \textbf{\bibinfo{volume}{95}}, \bibinfo{pages}{1642} (\bibinfo{year}{1988}).

\bibitem[{\citenamefont{{Gerbal} et~al.}(1992)\citenamefont{{Gerbal}, {Durret}, {Lachieze-Rey}, and {Lima-Neto}}}]{Gerbel}
\bibinfo{author}{\bibfnamefont{D.}~\bibnamefont{{Gerbal}}}, \bibinfo{author}{\bibfnamefont{F.}~\bibnamefont{{Durret}}}, \bibinfo{author}{\bibfnamefont{M.}~\bibnamefont{{Lachieze-Rey}}}, \bibnamefont{and} \bibinfo{author}{\bibfnamefont{G.}~\bibnamefont{{Lima-Neto}}}, \bibinfo{journal}{\aap} \textbf{\bibinfo{volume}{262}}, \bibinfo{pages}{395} (\bibinfo{year}{1992}).

\bibitem[{\citenamefont{{Aguirre} et~al.}(2001)\citenamefont{{Aguirre}, {Schaye}, and {Quataert}}}]{Aguirre}
\bibinfo{author}{\bibfnamefont{A.}~\bibnamefont{{Aguirre}}}, \bibinfo{author}{\bibfnamefont{J.}~\bibnamefont{{Schaye}}}, \bibnamefont{and} \bibinfo{author}{\bibfnamefont{E.}~\bibnamefont{{Quataert}}}, \bibinfo{journal}{\apj} \textbf{\bibinfo{volume}{561}}, \bibinfo{pages}{550} (\bibinfo{year}{2001}), \eprint{astro-ph/0105184}.

\bibitem[{\citenamefont{{Sanders}}(2003)}]{Sanders}
\bibinfo{author}{\bibfnamefont{R.~H.} \bibnamefont{{Sanders}}}, \bibinfo{journal}{\mnras} \textbf{\bibinfo{volume}{342}}, \bibinfo{pages}{901} (\bibinfo{year}{2003}), \eprint{astro-ph/0212293}.

\bibitem[{\citenamefont{{Angus} et~al.}(2008)\citenamefont{{Angus}, {Famaey}, and {Buote}}}]{Angus}
\bibinfo{author}{\bibfnamefont{G.~W.} \bibnamefont{{Angus}}}, \bibinfo{author}{\bibfnamefont{B.}~\bibnamefont{{Famaey}}}, \bibnamefont{and} \bibinfo{author}{\bibfnamefont{D.~A.} \bibnamefont{{Buote}}}, \bibinfo{journal}{\mnras} \textbf{\bibinfo{volume}{387}}, \bibinfo{pages}{1470} (\bibinfo{year}{2008}), \eprint{0709.0108}.

\bibitem[{\citenamefont{{Pointecouteau} and {Silk}}(2005)}]{Silk}
\bibinfo{author}{\bibfnamefont{E.}~\bibnamefont{{Pointecouteau}}} \bibnamefont{and} \bibinfo{author}{\bibfnamefont{J.}~\bibnamefont{{Silk}}}, \bibinfo{journal}{\mnras} \textbf{\bibinfo{volume}{364}}, \bibinfo{pages}{654} (\bibinfo{year}{2005}), \eprint{astro-ph/0505017}.

\bibitem[{\citenamefont{{Seeram} and {Desai}}(2021)}]{Seeram}
\bibinfo{author}{\bibfnamefont{S.}~\bibnamefont{{Seeram}}} \bibnamefont{and} \bibinfo{author}{\bibfnamefont{S.}~\bibnamefont{{Desai}}}, \bibinfo{journal}{Journal of Astrophysics and Astronomy} \textbf{\bibinfo{volume}{42}}, \bibinfo{eid}{3} (\bibinfo{year}{2021}).

\bibitem[{\citenamefont{{Reiprich} and {B{\"o}hringer}}(2002)}]{Reiprich}
\bibinfo{author}{\bibfnamefont{T.~H.} \bibnamefont{{Reiprich}}} \bibnamefont{and} \bibinfo{author}{\bibfnamefont{H.}~\bibnamefont{{B{\"o}hringer}}}, \bibinfo{journal}{\apj} \textbf{\bibinfo{volume}{567}}, \bibinfo{pages}{716} (\bibinfo{year}{2002}), \eprint{astro-ph/0111285}.

\bibitem[{\citenamefont{{Gastaldello} et~al.}(2014)\citenamefont{{Gastaldello}, {Limousin}, {Foex}, {Munoz}, {Verdugo}, {Motta}, {More}, {Cabanac}, {Buote}, {Eckert} et~al.}}]{Gasta}
\bibinfo{author}{\bibfnamefont{F.}~\bibnamefont{{Gastaldello}}}, \bibinfo{author}{\bibfnamefont{M.}~\bibnamefont{{Limousin}}}, \bibinfo{author}{\bibfnamefont{G.}~\bibnamefont{{Foex}}}, \bibinfo{author}{\bibfnamefont{R.~P.} \bibnamefont{{Munoz}}}, \bibinfo{author}{\bibfnamefont{T.}~\bibnamefont{{Verdugo}}}, \bibinfo{author}{\bibfnamefont{V.}~\bibnamefont{{Motta}}}, \bibinfo{author}{\bibfnamefont{A.}~\bibnamefont{{More}}}, \bibinfo{author}{\bibfnamefont{R.}~\bibnamefont{{Cabanac}}}, \bibinfo{author}{\bibfnamefont{D.~A.} \bibnamefont{{Buote}}}, \bibinfo{author}{\bibfnamefont{D.}~\bibnamefont{{Eckert}}}, \bibnamefont{et~al.}, \bibinfo{journal}{\mnras} \textbf{\bibinfo{volume}{442}}, \bibinfo{pages}{L76} (\bibinfo{year}{2014}), \eprint{1404.5633}.

\bibitem[{\citenamefont{{Natarajan} and {Zhao}}(2008)}]{Natarajan}
\bibinfo{author}{\bibfnamefont{P.}~\bibnamefont{{Natarajan}}} \bibnamefont{and} \bibinfo{author}{\bibfnamefont{H.}~\bibnamefont{{Zhao}}}, \bibinfo{journal}{\mnras} \textbf{\bibinfo{volume}{389}}, \bibinfo{pages}{250} (\bibinfo{year}{2008}), \eprint{0806.3080}.

\bibitem[{\citenamefont{{Takahashi} and {Chiba}}(2007)}]{Chiba}
\bibinfo{author}{\bibfnamefont{R.}~\bibnamefont{{Takahashi}}} \bibnamefont{and} \bibinfo{author}{\bibfnamefont{T.}~\bibnamefont{{Chiba}}}, \bibinfo{journal}{\apj} \textbf{\bibinfo{volume}{671}}, \bibinfo{pages}{45} (\bibinfo{year}{2007}), \eprint{astro-ph/0701365}.

\bibitem[{\citenamefont{{Gianfagna} et~al.}(2023)\citenamefont{{Gianfagna}, {Rasia}, {Cui}, {De Petris}, {Yepes}, {Contreras-Santos}, and {Knebe}}}]{massbias}
\bibinfo{author}{\bibfnamefont{G.}~\bibnamefont{{Gianfagna}}}, \bibinfo{author}{\bibfnamefont{E.}~\bibnamefont{{Rasia}}}, \bibinfo{author}{\bibfnamefont{W.}~\bibnamefont{{Cui}}}, \bibinfo{author}{\bibfnamefont{M.}~\bibnamefont{{De Petris}}}, \bibinfo{author}{\bibfnamefont{G.}~\bibnamefont{{Yepes}}}, \bibinfo{author}{\bibfnamefont{A.}~\bibnamefont{{Contreras-Santos}}}, \bibnamefont{and} \bibinfo{author}{\bibfnamefont{A.}~\bibnamefont{{Knebe}}}, \bibinfo{journal}{\mnras} \textbf{\bibinfo{volume}{518}}, \bibinfo{pages}{4238} (\bibinfo{year}{2023}), \eprint{2211.08372}.

\bibitem[{\citenamefont{{Tian} et~al.}(2020)\citenamefont{{Tian}, {Umetsu}, {Ko}, {Donahue}, and {Chiu}}}]{Tian}
\bibinfo{author}{\bibfnamefont{Y.}~\bibnamefont{{Tian}}}, \bibinfo{author}{\bibfnamefont{K.}~\bibnamefont{{Umetsu}}}, \bibinfo{author}{\bibfnamefont{C.-M.} \bibnamefont{{Ko}}}, \bibinfo{author}{\bibfnamefont{M.}~\bibnamefont{{Donahue}}}, \bibnamefont{and} \bibinfo{author}{\bibfnamefont{I.~N.} \bibnamefont{{Chiu}}}, \bibinfo{journal}{\apj} \textbf{\bibinfo{volume}{896}}, \bibinfo{eid}{70} (\bibinfo{year}{2020}), \eprint{2001.08340}.

\bibitem[{\citenamefont{{Chan} and {Del Popolo}}(2020)}]{Chan20}
\bibinfo{author}{\bibfnamefont{M.~H.} \bibnamefont{{Chan}}} \bibnamefont{and} \bibinfo{author}{\bibfnamefont{A.}~\bibnamefont{{Del Popolo}}}, \bibinfo{journal}{\mnras} p. \bibinfo{pages}{218} (\bibinfo{year}{2020}), \eprint{2001.06141}.

\bibitem[{\citenamefont{Pradyumna et~al.}(2021)\citenamefont{Pradyumna, Gupta, Seeram, and Desai}}]{pradyumna2021yet}
\bibinfo{author}{\bibfnamefont{S.}~\bibnamefont{Pradyumna}}, \bibinfo{author}{\bibfnamefont{S.}~\bibnamefont{Gupta}}, \bibinfo{author}{\bibfnamefont{S.}~\bibnamefont{Seeram}}, \bibnamefont{and} \bibinfo{author}{\bibfnamefont{S.}~\bibnamefont{Desai}}, \bibinfo{journal}{Physics of the Dark Universe} \textbf{\bibinfo{volume}{31}}, \bibinfo{pages}{100765} (\bibinfo{year}{2021}).

\bibitem[{\citenamefont{{Pradyumna} and {Desai}}(2021)}]{Pradyumna21}
\bibinfo{author}{\bibfnamefont{S.}~\bibnamefont{{Pradyumna}}} \bibnamefont{and} \bibinfo{author}{\bibfnamefont{S.}~\bibnamefont{{Desai}}}, \bibinfo{journal}{Physics of the Dark Universe} \textbf{\bibinfo{volume}{33}}, \bibinfo{eid}{100854} (\bibinfo{year}{2021}), \eprint{2107.05845}.

\bibitem[{\citenamefont{{Eckert} et~al.}(2022)\citenamefont{{Eckert}, {Ettori}, {Pointecouteau}, {van der Burg}, and {Loubser}}}]{Eckert22}
\bibinfo{author}{\bibfnamefont{D.}~\bibnamefont{{Eckert}}}, \bibinfo{author}{\bibfnamefont{S.}~\bibnamefont{{Ettori}}}, \bibinfo{author}{\bibfnamefont{E.}~\bibnamefont{{Pointecouteau}}}, \bibinfo{author}{\bibfnamefont{R.~F.~J.} \bibnamefont{{van der Burg}}}, \bibnamefont{and} \bibinfo{author}{\bibfnamefont{S.~I.} \bibnamefont{{Loubser}}}, \bibinfo{journal}{\aap} \textbf{\bibinfo{volume}{662}}, \bibinfo{eid}{A123} (\bibinfo{year}{2022}), \eprint{2205.01110}.

\bibitem[{\citenamefont{{Chan} and {Law}}(2022)}]{Chan22}
\bibinfo{author}{\bibfnamefont{M.~H.} \bibnamefont{{Chan}}} \bibnamefont{and} \bibinfo{author}{\bibfnamefont{K.~C.} \bibnamefont{{Law}}}, \bibinfo{journal}{\prd} \textbf{\bibinfo{volume}{105}}, \bibinfo{eid}{083003} (\bibinfo{year}{2022}), \eprint{2203.15217}.

\bibitem[{\citenamefont{{Chan}}(2014)}]{Chan14}
\bibinfo{author}{\bibfnamefont{M.~H.} \bibnamefont{{Chan}}}, \bibinfo{journal}{\mnras} \textbf{\bibinfo{volume}{442}}, \bibinfo{pages}{L14} (\bibinfo{year}{2014}), \eprint{1403.4352}.

\bibitem[{\citenamefont{{Gopika} and {Desai}}(2020)}]{Gopika}
\bibinfo{author}{\bibfnamefont{K.}~\bibnamefont{{Gopika}}} \bibnamefont{and} \bibinfo{author}{\bibfnamefont{S.}~\bibnamefont{{Desai}}}, \bibinfo{journal}{Physics of the Dark Universe} \textbf{\bibinfo{volume}{30}}, \bibinfo{eid}{100707} (\bibinfo{year}{2020}), \eprint{2006.12320}.

\bibitem[{\citenamefont{{Gopika} and {Desai}}(2021)}]{Gopika21}
\bibinfo{author}{\bibfnamefont{K.}~\bibnamefont{{Gopika}}} \bibnamefont{and} \bibinfo{author}{\bibfnamefont{S.}~\bibnamefont{{Desai}}}, \bibinfo{journal}{Physics of the Dark Universe} \textbf{\bibinfo{volume}{33}}, \bibinfo{eid}{100874} (\bibinfo{year}{2021}), \eprint{2106.07294}.

\bibitem[{\citenamefont{{Boran} et~al.}(2018)\citenamefont{{Boran}, {Desai}, {Kahya}, and {Woodard}}}]{Desai}
\bibinfo{author}{\bibfnamefont{S.}~\bibnamefont{{Boran}}}, \bibinfo{author}{\bibfnamefont{S.}~\bibnamefont{{Desai}}}, \bibinfo{author}{\bibfnamefont{E.~O.} \bibnamefont{{Kahya}}}, \bibnamefont{and} \bibinfo{author}{\bibfnamefont{R.~P.} \bibnamefont{{Woodard}}}, \bibinfo{journal}{\prd} \textbf{\bibinfo{volume}{97}}, \bibinfo{eid}{041501} (\bibinfo{year}{2018}), \eprint{1710.06168}.

\bibitem[{\citenamefont{{Ettori} et~al.}(2017)\citenamefont{{Ettori}, {Ghirardini}, {Eckert}, {Dubath}, and {Pointecouteau}}}]{Ettori16}
\bibinfo{author}{\bibfnamefont{S.}~\bibnamefont{{Ettori}}}, \bibinfo{author}{\bibfnamefont{V.}~\bibnamefont{{Ghirardini}}}, \bibinfo{author}{\bibfnamefont{D.}~\bibnamefont{{Eckert}}}, \bibinfo{author}{\bibfnamefont{F.}~\bibnamefont{{Dubath}}}, \bibnamefont{and} \bibinfo{author}{\bibfnamefont{E.}~\bibnamefont{{Pointecouteau}}}, \bibinfo{journal}{\mnras} \textbf{\bibinfo{volume}{470}}, \bibinfo{pages}{L29} (\bibinfo{year}{2017}), \eprint{1612.07288}.

\bibitem[{\citenamefont{{Ettori} et~al.}(2019)\citenamefont{{Ettori}, {Ghirardini}, {Eckert}, {Pointecouteau}, {Gastaldello}, {Sereno}, {Gaspari}, {Ghizzardi}, {Roncarelli}, and {Rossetti}}}]{Ettori18}
\bibinfo{author}{\bibfnamefont{S.}~\bibnamefont{{Ettori}}}, \bibinfo{author}{\bibfnamefont{V.}~\bibnamefont{{Ghirardini}}}, \bibinfo{author}{\bibfnamefont{D.}~\bibnamefont{{Eckert}}}, \bibinfo{author}{\bibfnamefont{E.}~\bibnamefont{{Pointecouteau}}}, \bibinfo{author}{\bibfnamefont{F.}~\bibnamefont{{Gastaldello}}}, \bibinfo{author}{\bibfnamefont{M.}~\bibnamefont{{Sereno}}}, \bibinfo{author}{\bibfnamefont{M.}~\bibnamefont{{Gaspari}}}, \bibinfo{author}{\bibfnamefont{S.}~\bibnamefont{{Ghizzardi}}}, \bibinfo{author}{\bibfnamefont{M.}~\bibnamefont{{Roncarelli}}}, \bibnamefont{and} \bibinfo{author}{\bibfnamefont{M.}~\bibnamefont{{Rossetti}}}, \bibinfo{journal}{\aap} \textbf{\bibinfo{volume}{621}}, \bibinfo{eid}{A39} (\bibinfo{year}{2019}), \eprint{1805.00035}.

\bibitem[{\citenamefont{{Halenka} and {Miller}}(2020)}]{Miller}
\bibinfo{author}{\bibfnamefont{V.}~\bibnamefont{{Halenka}}} \bibnamefont{and} \bibinfo{author}{\bibfnamefont{C.~J.} \bibnamefont{{Miller}}}, \bibinfo{journal}{\prd} \textbf{\bibinfo{volume}{102}}, \bibinfo{eid}{084007} (\bibinfo{year}{2020}), \eprint{1807.01689}.

\bibitem[{\citenamefont{{Tamosiunas} et~al.}(2019)\citenamefont{{Tamosiunas}, {Bacon}, {Koyama}, and {Nichol}}}]{Nichol}
\bibinfo{author}{\bibfnamefont{A.}~\bibnamefont{{Tamosiunas}}}, \bibinfo{author}{\bibfnamefont{D.}~\bibnamefont{{Bacon}}}, \bibinfo{author}{\bibfnamefont{K.}~\bibnamefont{{Koyama}}}, \bibnamefont{and} \bibinfo{author}{\bibfnamefont{R.~C.} \bibnamefont{{Nichol}}}, \bibinfo{journal}{\jcap} \textbf{\bibinfo{volume}{2019}}, \bibinfo{eid}{053} (\bibinfo{year}{2019}), \eprint{1901.05505}.

\bibitem[{\citenamefont{{ZuHone} and {Sims}}(2019)}]{ZuHone}
\bibinfo{author}{\bibfnamefont{J.~A.} \bibnamefont{{ZuHone}}} \bibnamefont{and} \bibinfo{author}{\bibfnamefont{J.}~\bibnamefont{{Sims}}}, \bibinfo{journal}{\apj} \textbf{\bibinfo{volume}{880}}, \bibinfo{eid}{145} (\bibinfo{year}{2019}), \eprint{1905.03832}.

\bibitem[{\citenamefont{{Pardo}}(2020)}]{Pardo}
\bibinfo{author}{\bibfnamefont{K.}~\bibnamefont{{Pardo}}}, \bibinfo{journal}{\jcap} \textbf{\bibinfo{volume}{2020}}, \bibinfo{eid}{012} (\bibinfo{year}{2020}), \eprint{1706.00785}.

\bibitem[{\citenamefont{{Diez-Tejedor} et~al.}(2018)\citenamefont{{Diez-Tejedor}, {Gonzalez-Morales}, and {Niz}}}]{Diez}
\bibinfo{author}{\bibfnamefont{A.}~\bibnamefont{{Diez-Tejedor}}}, \bibinfo{author}{\bibfnamefont{A.~X.} \bibnamefont{{Gonzalez-Morales}}}, \bibnamefont{and} \bibinfo{author}{\bibfnamefont{G.}~\bibnamefont{{Niz}}}, \bibinfo{journal}{\mnras} \textbf{\bibinfo{volume}{477}}, \bibinfo{pages}{1285} (\bibinfo{year}{2018}), \eprint{1612.06282}.

\bibitem[{\citenamefont{{Lelli} et~al.}(2017)\citenamefont{{Lelli}, {McGaugh}, and {Schombert}}}]{Lelli17}
\bibinfo{author}{\bibfnamefont{F.}~\bibnamefont{{Lelli}}}, \bibinfo{author}{\bibfnamefont{S.~S.} \bibnamefont{{McGaugh}}}, \bibnamefont{and} \bibinfo{author}{\bibfnamefont{J.~M.} \bibnamefont{{Schombert}}}, \bibinfo{journal}{\mnras} \textbf{\bibinfo{volume}{468}}, \bibinfo{pages}{L68} (\bibinfo{year}{2017}), \eprint{1702.04355}.

\bibitem[{\citenamefont{{Yoon} et~al.}(2023)\citenamefont{{Yoon}, {Park}, and {Hwang}}}]{Yoon}
\bibinfo{author}{\bibfnamefont{Y.}~\bibnamefont{{Yoon}}}, \bibinfo{author}{\bibfnamefont{J.-C.} \bibnamefont{{Park}}}, \bibnamefont{and} \bibinfo{author}{\bibfnamefont{H.~S.} \bibnamefont{{Hwang}}}, \bibinfo{journal}{Classical and Quantum Gravity} \textbf{\bibinfo{volume}{40}}, \bibinfo{eid}{02LT01} (\bibinfo{year}{2023}), \eprint{2206.11685}.

\bibitem[{\citenamefont{{Tortora} et~al.}(2018)\citenamefont{{Tortora}, {Koopmans}, {Napolitano}, and {Valentijn}}}]{Tortora}
\bibinfo{author}{\bibfnamefont{C.}~\bibnamefont{{Tortora}}}, \bibinfo{author}{\bibfnamefont{L.~V.~E.} \bibnamefont{{Koopmans}}}, \bibinfo{author}{\bibfnamefont{N.~R.} \bibnamefont{{Napolitano}}}, \bibnamefont{and} \bibinfo{author}{\bibfnamefont{E.~A.} \bibnamefont{{Valentijn}}}, \bibinfo{journal}{\mnras} \textbf{\bibinfo{volume}{473}}, \bibinfo{pages}{2324} (\bibinfo{year}{2018}), \eprint{1702.08865}.

\bibitem[{\citenamefont{{Brouwer} et~al.}(2017)\citenamefont{{Brouwer}, {Visser}, {Dvornik}, {Hoekstra}, {Kuijken}, {Valentijn}, {Bilicki}, {Blake}, {Brough}, {Buddelmeijer} et~al.}}]{Brouwer}
\bibinfo{author}{\bibfnamefont{M.~M.} \bibnamefont{{Brouwer}}}, \bibinfo{author}{\bibfnamefont{M.~R.} \bibnamefont{{Visser}}}, \bibinfo{author}{\bibfnamefont{A.}~\bibnamefont{{Dvornik}}}, \bibinfo{author}{\bibfnamefont{H.}~\bibnamefont{{Hoekstra}}}, \bibinfo{author}{\bibfnamefont{K.}~\bibnamefont{{Kuijken}}}, \bibinfo{author}{\bibfnamefont{E.~A.} \bibnamefont{{Valentijn}}}, \bibinfo{author}{\bibfnamefont{M.}~\bibnamefont{{Bilicki}}}, \bibinfo{author}{\bibfnamefont{C.}~\bibnamefont{{Blake}}}, \bibinfo{author}{\bibfnamefont{S.}~\bibnamefont{{Brough}}}, \bibinfo{author}{\bibfnamefont{H.}~\bibnamefont{{Buddelmeijer}}}, \bibnamefont{et~al.}, \bibinfo{journal}{\mnras} \textbf{\bibinfo{volume}{466}}, \bibinfo{pages}{2547} (\bibinfo{year}{2017}), \eprint{1612.03034}.

\bibitem[{\citenamefont{{Brouwer} et~al.}(2021)\citenamefont{{Brouwer}, {Oman}, {Valentijn}, {Bilicki}, {Heymans}, {Hoekstra}, {Napolitano}, {Roy}, {Tortora}, {Wright} et~al.}}]{Brouwer21}
\bibinfo{author}{\bibfnamefont{M.~M.} \bibnamefont{{Brouwer}}}, \bibinfo{author}{\bibfnamefont{K.~A.} \bibnamefont{{Oman}}}, \bibinfo{author}{\bibfnamefont{E.~A.} \bibnamefont{{Valentijn}}}, \bibinfo{author}{\bibfnamefont{M.}~\bibnamefont{{Bilicki}}}, \bibinfo{author}{\bibfnamefont{C.}~\bibnamefont{{Heymans}}}, \bibinfo{author}{\bibfnamefont{H.}~\bibnamefont{{Hoekstra}}}, \bibinfo{author}{\bibfnamefont{N.~R.} \bibnamefont{{Napolitano}}}, \bibinfo{author}{\bibfnamefont{N.}~\bibnamefont{{Roy}}}, \bibinfo{author}{\bibfnamefont{C.}~\bibnamefont{{Tortora}}}, \bibinfo{author}{\bibfnamefont{A.~H.} \bibnamefont{{Wright}}}, \bibnamefont{et~al.}, \bibinfo{journal}{\aap} \textbf{\bibinfo{volume}{650}}, \bibinfo{eid}{A113} (\bibinfo{year}{2021}), \eprint{2106.11677}.

\bibitem[{\citenamefont{{Luo} et~al.}(2021)\citenamefont{{Luo}, {Zhang}, {Halenka}, {Yang}, {More}, {Miller}, {Liu}, and {Shi}}}]{Luo21}
\bibinfo{author}{\bibfnamefont{W.}~\bibnamefont{{Luo}}}, \bibinfo{author}{\bibfnamefont{J.}~\bibnamefont{{Zhang}}}, \bibinfo{author}{\bibfnamefont{V.}~\bibnamefont{{Halenka}}}, \bibinfo{author}{\bibfnamefont{X.}~\bibnamefont{{Yang}}}, \bibinfo{author}{\bibfnamefont{S.}~\bibnamefont{{More}}}, \bibinfo{author}{\bibfnamefont{C.~J.} \bibnamefont{{Miller}}}, \bibinfo{author}{\bibfnamefont{L.}~\bibnamefont{{Liu}}}, \bibnamefont{and} \bibinfo{author}{\bibfnamefont{F.}~\bibnamefont{{Shi}}}, \bibinfo{journal}{\apj} \textbf{\bibinfo{volume}{914}}, \bibinfo{eid}{96} (\bibinfo{year}{2021}), \eprint{2003.09818}.

\bibitem[{\citenamefont{{Predehl} et~al.}(2021)\citenamefont{{Predehl}, {Andritschke}, {Arefiev}, {Babyshkin}, {Batanov}, {Becker}, {B{\"o}hringer}, {Bogomolov}, {Boller}, {Borm} et~al.}}]{Pre}
\bibinfo{author}{\bibfnamefont{P.}~\bibnamefont{{Predehl}}}, \bibinfo{author}{\bibfnamefont{R.}~\bibnamefont{{Andritschke}}}, \bibinfo{author}{\bibfnamefont{V.}~\bibnamefont{{Arefiev}}}, \bibinfo{author}{\bibfnamefont{V.}~\bibnamefont{{Babyshkin}}}, \bibinfo{author}{\bibfnamefont{O.}~\bibnamefont{{Batanov}}}, \bibinfo{author}{\bibfnamefont{W.}~\bibnamefont{{Becker}}}, \bibinfo{author}{\bibfnamefont{H.}~\bibnamefont{{B{\"o}hringer}}}, \bibinfo{author}{\bibfnamefont{A.}~\bibnamefont{{Bogomolov}}}, \bibinfo{author}{\bibfnamefont{T.}~\bibnamefont{{Boller}}}, \bibinfo{author}{\bibfnamefont{K.}~\bibnamefont{{Borm}}}, \bibnamefont{et~al.}, \bibinfo{journal}{\aap} \textbf{\bibinfo{volume}{647}}, \bibinfo{eid}{A1} (\bibinfo{year}{2021}), \eprint{2010.03477}.

\bibitem[{\citenamefont{{Bulbul} et~al.}(2022)\citenamefont{{Bulbul}, {Liu}, {Pasini}, {Comparat}, {Hoang}, {Klein}, {Ghirardini}, {Salvato}, {Merloni}, {Seppi} et~al.}}]{Bulbul21}
\bibinfo{author}{\bibfnamefont{E.}~\bibnamefont{{Bulbul}}}, \bibinfo{author}{\bibfnamefont{A.}~\bibnamefont{{Liu}}}, \bibinfo{author}{\bibfnamefont{T.}~\bibnamefont{{Pasini}}}, \bibinfo{author}{\bibfnamefont{J.}~\bibnamefont{{Comparat}}}, \bibinfo{author}{\bibfnamefont{D.~N.} \bibnamefont{{Hoang}}}, \bibinfo{author}{\bibfnamefont{M.}~\bibnamefont{{Klein}}}, \bibinfo{author}{\bibfnamefont{V.}~\bibnamefont{{Ghirardini}}}, \bibinfo{author}{\bibfnamefont{M.}~\bibnamefont{{Salvato}}}, \bibinfo{author}{\bibfnamefont{A.}~\bibnamefont{{Merloni}}}, \bibinfo{author}{\bibfnamefont{R.}~\bibnamefont{{Seppi}}}, \bibnamefont{et~al.}, \bibinfo{journal}{\aap} \textbf{\bibinfo{volume}{661}}, \bibinfo{eid}{A10} (\bibinfo{year}{2022}), \eprint{2110.09544}.

\bibitem[{\citenamefont{{Liu} et~al.}(2023)\citenamefont{{Liu}, {Bulbul}, {Ramos-Ceja}, {Sanders}, {Ghirardini}, {Bahar}, {Yeung}, {Gatuzz}, {Freyberg}, {Garrel} et~al.}}]{liu2022x}
\bibinfo{author}{\bibfnamefont{A.}~\bibnamefont{{Liu}}}, \bibinfo{author}{\bibfnamefont{E.}~\bibnamefont{{Bulbul}}}, \bibinfo{author}{\bibfnamefont{M.~E.} \bibnamefont{{Ramos-Ceja}}}, \bibinfo{author}{\bibfnamefont{J.~S.} \bibnamefont{{Sanders}}}, \bibinfo{author}{\bibfnamefont{V.}~\bibnamefont{{Ghirardini}}}, \bibinfo{author}{\bibfnamefont{Y.~E.} \bibnamefont{{Bahar}}}, \bibinfo{author}{\bibfnamefont{M.}~\bibnamefont{{Yeung}}}, \bibinfo{author}{\bibfnamefont{E.}~\bibnamefont{{Gatuzz}}}, \bibinfo{author}{\bibfnamefont{M.}~\bibnamefont{{Freyberg}}}, \bibinfo{author}{\bibfnamefont{C.}~\bibnamefont{{Garrel}}}, \bibnamefont{et~al.}, \bibinfo{journal}{\aap} \textbf{\bibinfo{volume}{670}}, \bibinfo{eid}{A96} (\bibinfo{year}{2023}), \eprint{2210.00633}.

\bibitem[{\citenamefont{{Caminha} et~al.}(2022)\citenamefont{{Caminha}, {Suyu}, {Mercurio}, {Brammer}, {Bergamini}, {Acebron}, and {Vanzella}}}]{Caminha22}
\bibinfo{author}{\bibfnamefont{G.~B.} \bibnamefont{{Caminha}}}, \bibinfo{author}{\bibfnamefont{S.~H.} \bibnamefont{{Suyu}}}, \bibinfo{author}{\bibfnamefont{A.}~\bibnamefont{{Mercurio}}}, \bibinfo{author}{\bibfnamefont{G.}~\bibnamefont{{Brammer}}}, \bibinfo{author}{\bibfnamefont{P.}~\bibnamefont{{Bergamini}}}, \bibinfo{author}{\bibfnamefont{A.}~\bibnamefont{{Acebron}}}, \bibnamefont{and} \bibinfo{author}{\bibfnamefont{E.}~\bibnamefont{{Vanzella}}}, \bibinfo{journal}{\aap} \textbf{\bibinfo{volume}{666}}, \bibinfo{eid}{L9} (\bibinfo{year}{2022}), \eprint{2207.07567}.

\bibitem[{\citenamefont{{Sarazin}}(1986)}]{Sarazin}
\bibinfo{author}{\bibfnamefont{C.~L.} \bibnamefont{{Sarazin}}}, \bibinfo{journal}{Reviews of Modern Physics} \textbf{\bibinfo{volume}{58}}, \bibinfo{pages}{1} (\bibinfo{year}{1986}).

\bibitem[{\citenamefont{Vikhlinin et~al.}(2006)\citenamefont{Vikhlinin, Kravtsov, Forman, Jones, Markevitch, Murray, and Van~Speybroeck}}]{vikhlinin2006chandra}
\bibinfo{author}{\bibfnamefont{A.}~\bibnamefont{Vikhlinin}}, \bibinfo{author}{\bibfnamefont{A.}~\bibnamefont{Kravtsov}}, \bibinfo{author}{\bibfnamefont{W.}~\bibnamefont{Forman}}, \bibinfo{author}{\bibfnamefont{C.}~\bibnamefont{Jones}}, \bibinfo{author}{\bibfnamefont{M.}~\bibnamefont{Markevitch}}, \bibinfo{author}{\bibfnamefont{S.}~\bibnamefont{Murray}}, \bibnamefont{and} \bibinfo{author}{\bibfnamefont{L.}~\bibnamefont{Van~Speybroeck}}, \bibinfo{journal}{The Astrophysical Journal} \textbf{\bibinfo{volume}{640}}, \bibinfo{pages}{691} (\bibinfo{year}{2006}).

\bibitem[{\citenamefont{{Cavaliere} and {Fusco-Femiano}}(1976)}]{betamodel}
\bibinfo{author}{\bibfnamefont{A.}~\bibnamefont{{Cavaliere}}} \bibnamefont{and} \bibinfo{author}{\bibfnamefont{R.}~\bibnamefont{{Fusco-Femiano}}}, \bibinfo{journal}{\aap} \textbf{\bibinfo{volume}{49}}, \bibinfo{pages}{137} (\bibinfo{year}{1976}).

\bibitem[{\citenamefont{{Upadhyaya} and {Desai}}(2023)}]{Varenya}
\bibinfo{author}{\bibfnamefont{V.}~\bibnamefont{{Upadhyaya}}} \bibnamefont{and} \bibinfo{author}{\bibfnamefont{S.}~\bibnamefont{{Desai}}}, \bibinfo{journal}{Physics of the Dark Universe} \textbf{\bibinfo{volume}{40}}, \bibinfo{eid}{101182} (\bibinfo{year}{2023}), \eprint{2212.02326}.

\bibitem[{\citenamefont{{Chiu} et~al.}(2018)\citenamefont{{Chiu}, {Mohr}, {McDonald}, {Bocquet}, {Desai}, {Klein}, {Israel}, {Ashby}, {Stanford}, {Benson} et~al.}}]{chiu2018baryon}
\bibinfo{author}{\bibfnamefont{I.}~\bibnamefont{{Chiu}}}, \bibinfo{author}{\bibfnamefont{J.~J.} \bibnamefont{{Mohr}}}, \bibinfo{author}{\bibfnamefont{M.}~\bibnamefont{{McDonald}}}, \bibinfo{author}{\bibfnamefont{S.}~\bibnamefont{{Bocquet}}}, \bibinfo{author}{\bibfnamefont{S.}~\bibnamefont{{Desai}}}, \bibinfo{author}{\bibfnamefont{M.}~\bibnamefont{{Klein}}}, \bibinfo{author}{\bibfnamefont{H.}~\bibnamefont{{Israel}}}, \bibinfo{author}{\bibfnamefont{M.~L.~N.} \bibnamefont{{Ashby}}}, \bibinfo{author}{\bibfnamefont{A.}~\bibnamefont{{Stanford}}}, \bibinfo{author}{\bibfnamefont{B.~A.} \bibnamefont{{Benson}}}, \bibnamefont{et~al.}, \bibinfo{journal}{\mnras} \textbf{\bibinfo{volume}{478}}, \bibinfo{pages}{3072} (\bibinfo{year}{2018}), \eprint{1711.00917}.

\bibitem[{\citenamefont{{Contini} et~al.}(2024)\citenamefont{{Contini}, {Yi}, and {Jeon}}}]{Contini}
\bibinfo{author}{\bibfnamefont{E.}~\bibnamefont{{Contini}}}, \bibinfo{author}{\bibfnamefont{S.~K.} \bibnamefont{{Yi}}}, \bibnamefont{and} \bibinfo{author}{\bibfnamefont{S.}~\bibnamefont{{Jeon}}}, \bibinfo{journal}{arXiv e-prints} \bibinfo{eid}{arXiv:2404.01560} (\bibinfo{year}{2024}), \eprint{2404.01560}.

\bibitem[{\citenamefont{{Mahler} et~al.}(2023)\citenamefont{{Mahler}, {Jauzac}, {Richard}, {Beauchesne}, {Ebeling}, {Lagattuta}, {Natarajan}, {Sharon}, {Atek}, {Claeyssens} et~al.}}]{Mahler}
\bibinfo{author}{\bibfnamefont{G.}~\bibnamefont{{Mahler}}}, \bibinfo{author}{\bibfnamefont{M.}~\bibnamefont{{Jauzac}}}, \bibinfo{author}{\bibfnamefont{J.}~\bibnamefont{{Richard}}}, \bibinfo{author}{\bibfnamefont{B.}~\bibnamefont{{Beauchesne}}}, \bibinfo{author}{\bibfnamefont{H.}~\bibnamefont{{Ebeling}}}, \bibinfo{author}{\bibfnamefont{D.}~\bibnamefont{{Lagattuta}}}, \bibinfo{author}{\bibfnamefont{P.}~\bibnamefont{{Natarajan}}}, \bibinfo{author}{\bibfnamefont{K.}~\bibnamefont{{Sharon}}}, \bibinfo{author}{\bibfnamefont{H.}~\bibnamefont{{Atek}}}, \bibinfo{author}{\bibfnamefont{A.}~\bibnamefont{{Claeyssens}}}, \bibnamefont{et~al.}, \bibinfo{journal}{\apj} \textbf{\bibinfo{volume}{945}}, \bibinfo{eid}{49} (\bibinfo{year}{2023}), \eprint{2207.07101}.

\bibitem[{\citenamefont{{Pascale} et~al.}(2022)\citenamefont{{Pascale}, {Frye}, {Diego}, {Furtak}, {Zitrin}, {Broadhurst}, {Conselice}, {Dai}, {Ferreira}, {Adams} et~al.}}]{Pascale}
\bibinfo{author}{\bibfnamefont{M.}~\bibnamefont{{Pascale}}}, \bibinfo{author}{\bibfnamefont{B.~L.} \bibnamefont{{Frye}}}, \bibinfo{author}{\bibfnamefont{J.}~\bibnamefont{{Diego}}}, \bibinfo{author}{\bibfnamefont{L.~J.} \bibnamefont{{Furtak}}}, \bibinfo{author}{\bibfnamefont{A.}~\bibnamefont{{Zitrin}}}, \bibinfo{author}{\bibfnamefont{T.}~\bibnamefont{{Broadhurst}}}, \bibinfo{author}{\bibfnamefont{C.~J.} \bibnamefont{{Conselice}}}, \bibinfo{author}{\bibfnamefont{L.}~\bibnamefont{{Dai}}}, \bibinfo{author}{\bibfnamefont{L.}~\bibnamefont{{Ferreira}}}, \bibinfo{author}{\bibfnamefont{N.~J.} \bibnamefont{{Adams}}}, \bibnamefont{et~al.}, \bibinfo{journal}{\apjl} \textbf{\bibinfo{volume}{938}}, \bibinfo{eid}{L6} (\bibinfo{year}{2022}), \eprint{2207.07102}.

\bibitem[{\citenamefont{{Bellstedt} et~al.}(2016)\citenamefont{{Bellstedt}, {Lidman}, {Muzzin}, {Franx}, {Guatelli}, {Hill}, {Hoekstra}, {Kurinsky}, {Labbe}, {Marchesini} et~al.}}]{Bellstedt}
\bibinfo{author}{\bibfnamefont{S.}~\bibnamefont{{Bellstedt}}}, \bibinfo{author}{\bibfnamefont{C.}~\bibnamefont{{Lidman}}}, \bibinfo{author}{\bibfnamefont{A.}~\bibnamefont{{Muzzin}}}, \bibinfo{author}{\bibfnamefont{M.}~\bibnamefont{{Franx}}}, \bibinfo{author}{\bibfnamefont{S.}~\bibnamefont{{Guatelli}}}, \bibinfo{author}{\bibfnamefont{A.~R.} \bibnamefont{{Hill}}}, \bibinfo{author}{\bibfnamefont{H.}~\bibnamefont{{Hoekstra}}}, \bibinfo{author}{\bibfnamefont{N.}~\bibnamefont{{Kurinsky}}}, \bibinfo{author}{\bibfnamefont{I.}~\bibnamefont{{Labbe}}}, \bibinfo{author}{\bibfnamefont{D.}~\bibnamefont{{Marchesini}}}, \bibnamefont{et~al.}, \bibinfo{journal}{\mnras} \textbf{\bibinfo{volume}{460}}, \bibinfo{pages}{2862} (\bibinfo{year}{2016}), \eprint{1605.02736}.

\bibitem[{\citenamefont{{Woodard}}(2015)}]{Woodard14}
\bibinfo{author}{\bibfnamefont{R.~P.} \bibnamefont{{Woodard}}}, \bibinfo{journal}{Canadian Journal of Physics} \textbf{\bibinfo{volume}{93}}, \bibinfo{pages}{242} (\bibinfo{year}{2015}), \eprint{1403.6763}.

\bibitem[{\citenamefont{{Kim} et~al.}(2016)\citenamefont{{Kim}, {Rahat}, {Sayeb}, {Tan}, {Woodard}, and {Xu}}}]{Woodard16}
\bibinfo{author}{\bibfnamefont{M.}~\bibnamefont{{Kim}}}, \bibinfo{author}{\bibfnamefont{M.~H.} \bibnamefont{{Rahat}}}, \bibinfo{author}{\bibfnamefont{M.}~\bibnamefont{{Sayeb}}}, \bibinfo{author}{\bibfnamefont{L.}~\bibnamefont{{Tan}}}, \bibinfo{author}{\bibfnamefont{R.~P.} \bibnamefont{{Woodard}}}, \bibnamefont{and} \bibinfo{author}{\bibfnamefont{B.}~\bibnamefont{{Xu}}}, \bibinfo{journal}{\prd} \textbf{\bibinfo{volume}{94}}, \bibinfo{eid}{104009} (\bibinfo{year}{2016}), \eprint{1608.07858}.

\bibitem[{\citenamefont{{Skordis} and {Z{\l}o{\'s}nik}}(2021)}]{Skordis}
\bibinfo{author}{\bibfnamefont{C.}~\bibnamefont{{Skordis}}} \bibnamefont{and} \bibinfo{author}{\bibfnamefont{T.}~\bibnamefont{{Z{\l}o{\'s}nik}}}, \bibinfo{journal}{\prl} \textbf{\bibinfo{volume}{127}}, \bibinfo{eid}{161302} (\bibinfo{year}{2021}), \eprint{2007.00082}.

\bibitem[{\citenamefont{{Hossenfelder}}(2017)}]{Sabine}
\bibinfo{author}{\bibfnamefont{S.}~\bibnamefont{{Hossenfelder}}}, \bibinfo{journal}{\prd} \textbf{\bibinfo{volume}{95}}, \bibinfo{eid}{124018} (\bibinfo{year}{2017}), \eprint{1703.01415}.

\bibitem[{\citenamefont{{Barret} et~al.}(2020)\citenamefont{{Barret}, {Decourchelle}, {Fabian}, {Guainazzi}, {Nandra}, {Smith}, and {den Herder}}}]{Athena}
\bibinfo{author}{\bibfnamefont{D.}~\bibnamefont{{Barret}}}, \bibinfo{author}{\bibfnamefont{A.}~\bibnamefont{{Decourchelle}}}, \bibinfo{author}{\bibfnamefont{A.}~\bibnamefont{{Fabian}}}, \bibinfo{author}{\bibfnamefont{M.}~\bibnamefont{{Guainazzi}}}, \bibinfo{author}{\bibfnamefont{K.}~\bibnamefont{{Nandra}}}, \bibinfo{author}{\bibfnamefont{R.}~\bibnamefont{{Smith}}}, \bibnamefont{and} \bibinfo{author}{\bibfnamefont{J.-W.} \bibnamefont{{den Herder}}}, \bibinfo{journal}{Astronomische Nachrichten} \textbf{\bibinfo{volume}{341}}, \bibinfo{pages}{224} (\bibinfo{year}{2020}), \eprint{1912.04615}.

\bibitem[{\citenamefont{{Zhang} et~al.}(2019)\citenamefont{{Zhang}, {Santangelo}, {Feroci}, {Xu}, {Lu}, {Chen}, {Feng}, {Zhang}, {Brandt}, {Hernanz} et~al.}}]{eXTP}
\bibinfo{author}{\bibfnamefont{S.}~\bibnamefont{{Zhang}}}, \bibinfo{author}{\bibfnamefont{A.}~\bibnamefont{{Santangelo}}}, \bibinfo{author}{\bibfnamefont{M.}~\bibnamefont{{Feroci}}}, \bibinfo{author}{\bibfnamefont{Y.}~\bibnamefont{{Xu}}}, \bibinfo{author}{\bibfnamefont{F.}~\bibnamefont{{Lu}}}, \bibinfo{author}{\bibfnamefont{Y.}~\bibnamefont{{Chen}}}, \bibinfo{author}{\bibfnamefont{H.}~\bibnamefont{{Feng}}}, \bibinfo{author}{\bibfnamefont{S.}~\bibnamefont{{Zhang}}}, \bibinfo{author}{\bibfnamefont{S.}~\bibnamefont{{Brandt}}}, \bibinfo{author}{\bibfnamefont{M.}~\bibnamefont{{Hernanz}}}, \bibnamefont{et~al.}, \bibinfo{journal}{Science China Physics, Mechanics, and Astronomy} \textbf{\bibinfo{volume}{62}}, \bibinfo{eid}{29502} (\bibinfo{year}{2019}), \eprint{1812.04020}.

\end{thebibliography}
\end{document}